\documentclass[conference]{IEEEtran}

\usepackage{cite}
\usepackage{amsmath,amssymb,amsfonts}
\usepackage{algorithmic}
\usepackage{graphicx}
\usepackage{textcomp}
\usepackage{hyperref}
\usepackage{cleveref}
\usepackage[table]{xcolor}
\usepackage{multirow}
\usepackage{booktabs}
\usepackage{arydshln}
\usepackage{colortbl}
\def\BibTeX{{\rm B\kern-.05em{\sc i\kern-.025em b}\kern-.08em
    T\kern-.1667em\lower.7ex\hbox{E}\kern-.125emX}}

\begin{document}

\title{Hierarchical Label Propagation: \\
A Model-Size-Dependent Performance Booster for AudioSet Tagging
}

\author{\IEEEauthorblockN{Ludovic Tuncay, Etienne Labbé and Thomas Pellegrini}
\IEEEauthorblockA{\textit{IRIT, Université de Toulouse, CNRS, Toulouse INP}\\
Toulouse, France \\
\{ludovic.tuncay, etienne.labbe, thomas.pellegrini\}@irit.fr}}

\hyphenation{PaSST-B ConvNeXt AudioSet PANN}

\maketitle

\begin{abstract}
AudioSet is one of the most used and largest datasets in audio tagging, containing about 2 million audio samples that are manually labeled with 527 event categories organized into an ontology. However, the annotations contain inconsistencies, particularly where categories that should be labeled as positive according to the ontology are frequently mislabeled as negative.
To address this issue, we apply Hierarchical Label Propagation (HLP), which propagates labels up the ontology hierarchy, resulting in a mean increase in positive labels per audio clip from 1.98 to 2.39 and affecting 109 out of the 527 classes.
Our results demonstrate that HLP provides performance benefits across various model architectures, including convolutional neural networks (PANN's CNN6 and ConvNeXT) and transformers (PaSST), with smaller models showing more improvements. 
Finally, on FSD50K, another widely used dataset, models trained on AudioSet with HLP consistently outperformed those trained without HLP. Our source code will be made available on GitHub.
\end{abstract}


\section{Introduction}
\label{sec:intro}
In machine learning and audio processing, data quality is crucial for model performance~\cite{Jain2020DataQuality, northcutt2021pervasive}. In this work, we address the problem of hierarchical label inconsistency in AudioSet~\cite{audioset}, one of the largest and most widely used datasets in multi-label audio tagging. While AudioSet provides an unprecedented scale and diversity of audio samples, its annotations contain inconsistencies, particularly where categories that should be labeled as positive according to the ontology are frequently mislabeled as negative.

A number of studies have already explored the benefits of exploiting label categories organized into an ontology, either to inform a model during training, or to tackle label quality issues. Early work by Lo et al.~\cite{Lo2011Cost-Sensitive} recognized the importance of label correlations in audio tagging, proposing a cost-sensitive multi-label learning approach. Building on this, graph-based methods emerged as powerful tools for modeling label dependencies. Wang et al.~\cite{Wang2020Modeling} introduced a method using Graph Convolutional Networks (GCN) to capture label dependencies, achieving significant improvements in mean average precision on AudioSet. Extending this paradigm, Shrivastava et al.~\cite{9053065} developed MT-GCN, a Multi-task Learning based Graph Convolutional Network that learns domain knowledge from ontology, effectively improving model robustness against label noise.

Recent innovations have focused on mitigating the impact of noisy labels in large audio datasets. Zhu et al.~\cite{Zhu2020Audio} introduced CrossFilter, a framework employing incremental filtering and multi-task learning, demonstrating significant performance improvements on datasets such as FSDKaggle2018~\cite{fonseca2018generalpurposetaggingfreesoundaudio} and FSDKaggle2019~\cite{fonseca2020audiotaggingnoisylabels}. Similarly, Gong et al.~\cite{Gong2021PSLA} proposed PSLA (Pretraining, Sampling, Labeling, and Aggregation), a collection of training techniques including label enhancement, achieving competitive results on AudioSet and state-of-the-art performance on FSD50K~\cite{fonseca2022fsd50kopendatasethumanlabeled}, another widely used audio tagging dataset.

Hierarchical Label Propagation (HLP) has emerged as a potential technique to leverage label relationships~\cite{fonseca2022fsd50kopendatasethumanlabeled}, yet its effectiveness remains underexplored. Our study addresses this gap by examining HLP's impact on a spectrum of model sizes and architectures for AudioSet tagging. We hypothesize that HLP's benefits may vary with model size, potentially offering more significant improvements to smaller models.

By implementing HLP on the AudioSet ontology, and evaluating it on various model sizes using mean Average Precision (mAP), we seek to inform more efficient strategies for audio tagging. Unlike architecture-specific methods, our HLP approach is universally applicable to any model predicting on ontologies. By examining the interplay between HLP and model size, we aim to provide a more comprehensive understanding of how these factors influence tagging performance. The insights gained from this study could have broader implications, potentially informing methods to ensure data integrity or increase annotation efficiency and density in hierarchical multi-label datasets. 

Finally, to facilitate reproducibility and future research, we make our implementation publicly available at \url{https://github.com/LudovicTuncay/hierarchical-label-propagation}.

\section{Methodology}

\subsection{Hierarchical Label Propagation}

\begin{figure}[t!]
    \centering
    
    \includegraphics[width=\linewidth]{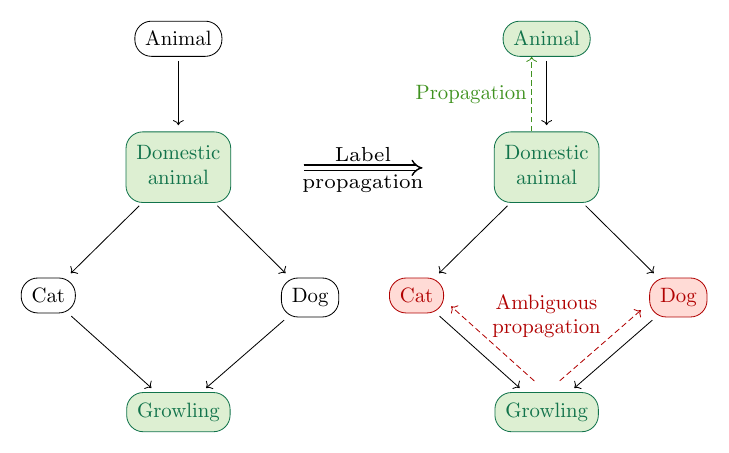}
    \caption{Illustration of HLP in a taxonomic ontology. The left side represents the initial state with \textit{Domestic animal} and \textit{Growling} as positive labels (green nodes). The right side depicts the state after applying HLP, where the positive label is propagated upwards to \textit{Animal} (green dashed arrow). The \textit{Cat} and \textit{Dog} nodes demonstrate potential ambiguous propagation (red nodes and dashed arrows), illustrating a challenge in HLP where the presence of \textit{Growling} does not uniquely determine its source between sibling nodes. In our method, these ambiguous labels are kept as negatives.}
    \label{fig:HLP}
    
\end{figure}

HLP is a technique that leverages the inherent structure of label ontologies to enhance multi-label classification. While applicable to any hierarchical label structure, our implementation utilizes the AudioSet ontology, propagating labels from specific classes to their parent classes.

As illustrated in Fig. \ref{fig:HLP}, HLP propagates positive labels upwards through the ontology. For instance, if \textit{Domestic animal} is labeled positive, the label is propagated to its parent node \textit{Animal}. This pre-processing step enriches the label set, providing the model with more comprehensive information during training.

The application of HLP serves several purposes: it enforces consistency in the label space, potentially improves the model's ability to recognize relationships between specific and general categories, and may help in scenarios where specific labels are sparse by providing more instances of general categories for training.

However, HLP introduces complexities, particularly when dealing with sibling nodes. As shown in Fig. \ref{fig:HLP}, the presence of \textit{Growling} creates ambiguity for its parent nodes \textit{Cat} and \textit{Dog}. We decide to keep these ambiguous nodes as negatives to avoid propagating uncertain and potentially incorrect labels through the hierarchy. This limitation highlights that HLP cannot resolve uncertainties between sibling nodes at the same hierarchical level. Additionally, if original labels are noisy or incorrect, HLP would propagate these errors upwards in the hierarchy.

In our implementation for AudioSet tagging, we apply HLP either as a pre-processing step to training labels or as a post-processing step to a model's outputs. As with pre-processing's single-parent constraint, post-processing only propagates when a child node $c$ has a single parent $p$, updating the parent's logit score $s_p = \max(s_p, s_c)$ if the child's logit score $s_c$ is greater.

\subsection{Model Architectures}

To investigate the impact of HLP across different model sizes, we selected four architectures varying in complexity: CNN6 from PANN \cite{kong2020pannslargescalepretrainedaudio}, ConvNeXt nano and femto \cite{liu2022convnet2020s, pellegrini2023adaptingconvnextmodelaudio}, and PaSST-B \cite{koutini22_interspeech}. Table \ref{tab:model_params} summarizes the parameter counts of these models. We chose not to keep the ConvNeXt-tiny architecture used in \cite{pellegrini2023adaptingconvnextmodelaudio}, as it was too demanding in training time and resources. 

This range of models, from 4.8M to 86.1M parameters, enables a comprehensive analysis of HLP's effectiveness across various model capacities in the context of AudioSet tagging.
\begin{table}[t!]
    \centering
    
    \caption{Summary of model architectures studied.}
    
    \label{tab:model_params}
    \begin{tabular}{lcc}
        \hline
        \textbf{Model} & \textbf{Parameters} & \textbf{Architecture type}\\
        \hline
        CNN6 (PANN) & \hphantom{1}4.8M & CNN\\
        ConvNeXt-femto & \hphantom{1}5.0M & CNN\\
        ConvNeXt-nano & 15.5M & CNN\\
        PaSST-B & 86.1M & Transformer\\
        \hline
    \end{tabular}
\end{table}

\subsection{Datasets}
Our study primarily utilizes AudioSet \cite{audioset}, a large-scale dataset of manually annotated audio events. It originally comprised over 2 million 10-second sound clips extracted from YouTube videos, spanning 527 classes organized in a hierarchical ontology of 632 classes. In our version of the dataset, we have 19,393 examples in the validation split, 21,022 in the ``balanced" training split, and 1,900,960 in the ``unbalanced" training split. For our experiments, we combine the balanced and unbalanced splits to form our training set. It is worth noting that since AudioSet's clips are sourced from YouTube, the dataset's composition may vary when downloaded at different times due to video edits or removals. This variability can complicate model performance comparisons. To mitigate this issue, we retrain all tested models on our specific version of the dataset. AudioSet's multi-label nature and diverse sound sources make it a valuable benchmark for audio event recognition tasks, despite the challenge of noisy labels \cite{audioset, northcutt2021pervasive}.

We also incorporate FSD50K \cite{fonseca2022fsd50kopendatasethumanlabeled} as additional evaluation data. This dataset contains 51,197 audio clips annotated with 200 classes from the AudioSet ontology, offering higher label quality and using Creative Commons licensed audio. While we train our models on AudioSet due to its larger scale and comprehensive ontology, we evaluate them on both AudioSet and FSD50K without any fine-tuning. This dual-dataset approach helps mitigate potential biases and label noise in AudioSet while validating our HLP technique. By evaluating HLP-trained models on both datasets without fine-tuning, we aim to isolate and assess the impact of our HLP technique on model performance. This methodology allows us to directly compare models trained with and without HLP, providing insights into the effectiveness of HLP across different data sources. It is important to note that we only use 192 classes in our evaluation on FSD50K, as these are the classes shared between AudioSet and FSD50K, ensuring consistency in our cross-dataset evaluation without fine-tuning.

While our experiments focus on AudioSet and FSD50K, it is important to note that HLP can be applied to any dataset with a hierarchical ontology structure, as the method only requires well-defined parent-child relationships between labels. This makes it a valuable tool for improving label consistency across various domains and datasets.

\subsection{Experimental Setup and Training Procedure}

Our experimental setup was designed to rigorously compare the performance of models trained with and without HLP. Experiments were conducted on the Jean-Zay supercomputer, utilizing eight V100 GPUs for smaller models (CNN6, ConvNeXt femto), four A100 GPUs for ConvNeXt nano, and a single RTX 4070 Ti SUPER for PaSST-B due to its faster training time.

We retrained every model using our curated version of the AudioSet dataset. HLP was optionally applied as a pre-processing step during data loading, modifying the original labels of the AudioSet dataset. For each model, all hyperparameters were shared between its HLP and non-HLP variants, ensuring that the application of HLP was the only difference between experimental conditions. Additionally, in specific instances, we applied HLP as a post-processing step to the model outputs to further examine its effects.

For all models, we adhered to the training instructions provided by the original authors, which included the use of label smoothing~\cite{szegedy2016rethinking}, and made minor adaptations to accommodate our experimental setup. No specific sampler was used for most models, with the exception of PaSST-B, which utilized the sampler implemented in the original authors' code. It is worth noting that HLP was not applied to FSD50K during our experiments, as this dataset already incorporates label propagation as part of its curation process.

Our evaluation methodology was comprehensive and multi-faceted. We first verified that the mAP improved when evaluating on an HLP-processed validation set, indicating better alignment between HLP datasets and model predictions. We then compared mAP between the HLP and non-HLP variants of each model on both AudioSet and FSD50K. 

By evaluating HLP-trained models on AudioSet and FSD50K without fine-tuning, we aim to assess the generalizability of our HLP technique. This approach enables a fair comparison between models trained with and without HLP, despite the potential for lower absolute performance on FSD50K when compared to models finetuned on this dataset.

\section{Results}

\subsection{Effects on Labels}
\begin{table}[t]
    
    \centering
    \caption{Summary of label changes due to HLP.}
    
    \label{tab:label_changes}
    \begin{tabular}{lcc}
        \hline
        \textbf{Metric} & \textbf{Before HLP} & \textbf{After HLP} \\
        \hline
        Affected classes & N/A & 109 \\
        Affected audios & N/A & 513,773\\
        Total labels & 3.85M & 4.64M\\
        Avg. labels / clip & 1.98 & 2.39\\
        \hline
    \end{tabular}
    
\end{table}

HLP significantly altered the label distribution in the AudioSet dataset. Table \ref{tab:label_changes} summarizes these changes.

HLP increased the total number of labels from 3.85M to 4.64M (+20.6\%), affecting 109 out of 527 classes (21\%) and more than 500K out of 2M audios (26\%). The average number of positive labels per audio clip rose from 1.98 to 2.39 (+20.6\%), indicating a denser label space. Some classes experienced substantial growth in occurrences. For example, \textit{Wild animals} grew from 1,112 to 40,505 occurrences (×36.4), \textit{Alarm} went from 726 to 15,603 occurrences (×21.5), and \textit{Motor vehicle (road)} increased from 8,954 to 71,684 occurrences (×8.0). Conversely, some specific classes remained largely unchanged: \textit{Speech} changed from 1,005,156 to 1,009,957 occurrences (×1.00), and \textit{Siren} moved from 8,286 to 8,518 occurrences (×1.03). These changes demonstrate HLP's impact on label distribution, particularly for higher-level categories in the AudioSet ontology.

\subsection{Effects on Model Performance}
\begin{table*}[t]
    \centering
    \caption{mAP comparison of models trained with and without HLP on various test sets. Bold values indicate the best performing configuration for each model. AS denotes the AudioSet dataset, AS-HLP refers to AudioSet with HLP applied during training, and AS-HLP~+~post HLP indicates evaluation with HLP applied to model predictions. mAP values are reported as percentages~(\%). $^\dagger$ denotes mAP \textbf{with} finetuning on FSD50K while $^*$ is mAP \textbf{without} finetuning.}
    \label{tab:model_performance}
    \begin{tabular}{lcccccc}
        \toprule
        \multirow{2}[2]{*}{Model} & \multirow{2}[2]{*}{\#Params (M)} & \multirow{2}[2]{*}{Train data} & \multicolumn{4}{c}{Test data} \\
        \cmidrule(lr){4-7}
        & & & AS-Base & AS-HLP & AS-HLP + post HLP & FSD50K \\
        \midrule
        CED-Small~\cite{dinkel2023ced} & 22 & AS-Base &  49.6 & N/A & N/A &  64.3$^\dagger$\\
        ATST-C2F~\cite{LI_2022} & 86 & AS-Base & 49.7 & N/A & N/A & 65.5$^\dagger$\\
        PaSST-B~\cite{koutini22_interspeech} & 86.1 & AS-Base & 46.2 & N/A & N/A & 64.9$^\dagger$ \\
        \noalign{\vskip 0.5mm}\hdashline\noalign{\vskip 0.5mm}
        &  & AS-Base & \textbf{30.6} & 31.1 & 32.8 & 31.3$^*$ \\
        \multirow{-2}{*}{CNN6} & \multirow{-2}{*}{4.8} & AS-HLP & 30.4 & \textbf{34.2} & \textbf{34.3} & \textbf{34.3}$^*$ \\
        \arrayrulecolor{gray!25}\midrule
        & & AS-Base & \textbf{37.4} & 37.7 & 37.6 & 33.4$^*$ \\
        \multirow{-2}{*}{ConvNeXt-femto} & \multirow{-2}{*}{5.0} & AS-HLP & 36.4 & \textbf{38.6} & \textbf{38.5} & \textbf{34.6}$^*$ \\
        \arrayrulecolor{gray!25}\midrule
        &  & AS-Base & \textbf{40.1} & \textbf{41.4} & \textbf{41.4} & 42.3$^*$ \\
        \multirow{-2}{*}{ConvNeXt-nano} & \multirow{-2}{*}{15.5} & AS-HLP & 39.9 & 41.2 & 41.0 & \textbf{42.4}$^*$ \\
        \arrayrulecolor{gray!25}\midrule
         & & AS-Base & \textbf{45.0} & 47.5 & 47.6 & 51.0$^*$\\
        \multirow{-2}{*}{PaSST-B} & \multirow{-2}{*}{86.1} & AS-HLP & 44.9 & \textbf{47.8} & \textbf{47.8} & \textbf{51.6}$^*$  \\
        \arrayrulecolor{black}
        \bottomrule
    \end{tabular}
    \vspace{-0.5em}
\end{table*}

\Cref{tab:model_performance} presents the mAP in percentage for various models trained and evaluated under different conditions. The results reveal intriguing patterns in the effectiveness of HLP across model sizes and evaluation scenarios.

Examining the performance of smaller models (CNN6 and ConvNeXt-femto), we observe a clear benefit from training with HLP when evaluated on AudioSet with HLP applied. For instance, CNN6 shows an improvement from 31.1\% to 34.2\% mAP, while ConvNeXt-femto improves from 37.7\% to 38.6\%. This supports our hypothesis that HLP aids smaller models in learning the ontology structure by reducing noise in label relationships. However, this advantage results in decreased performance when evaluated on the base AudioSet, where these models show slightly lower performance (e.g., CNN6 drops from 30.6\% to 30.4\%). This suggests that smaller models trained with HLP learn to respect the ontology structure, which may not align perfectly with the non-HLP test set.

In contrast, larger models (ConvNeXt-nano and PaSST-B) show less pronounced benefits from HLP training. Their performance remains relatively stable across different training and evaluation scenarios. For example, PaSST-B's performance on the HLP-processed AudioSet validation set remains nearly constant whether trained on the original AudioSet (47.5\% mAP) or with HLP-enhanced training data (47.8\% mAP). This suggests that larger models may already internalize the ontology structure during training, even without explicit HLP.

Applying HLP as a post-processing step on model outputs (AS-HLP + post HLP column) further highlights the size-dependent effects. Smaller models not trained with HLP see additional improvements (e.g., CNN6 from 31.1\% to 32.8\%), while models trained with HLP and larger models show minimal to no changes. This indicates that enforcing ontology hierarchy on outputs primarily benefits models that haven't fully learned the structure during training.

Evaluating on FSD50K provides additional insights. Smaller models trained with HLP show clear improvements (e.g., CNN6 from 31.3\% to 34.3\%), while larger models exhibit only slight to no advantages (e.g., ConvNeXt-nano from 42.3\% to 42.4\%). This consistent pattern across datasets reinforces our findings on the size-dependent benefits of HLP.

It is worth noting that the slight decreases in scores for some models when applying HLP as post-processing (e.g., ConvNeXt-nano from 41.2\% to 41.0\%) can be attributed to the propagation of erroneous predictions. This highlights the need to balance enforcing ontological structure with potential drawbacks of propagating errors.

\section{Conclusion} 

In this paper, we studied Hierarchical Label Propagation (HLP) which removes hierarchical inconsistencies in the label space. We applied this strategy on AudioSet and FSD50K either through training or evaluation.
HLP increased the mean number of positive labels per audio from 1.98 to 2.39 in AudioSet. This led to improved model performance across architectures and datasets, particularly benefiting smaller models. For instance, the CNN6 model saw a 3.1 percentage points increase in mAP on AudioSet with HLP, while larger models like PaSST-B showed modest gains of 0.3 percentage point, suggesting they may already better internalize the ontology structure. Similar findings were observed in the FSD50K dataset when using models trained on AudioSet directly to perform tagging on FSD50K.

While promising, our approach faces limitations with noisy labels, as HLP could propagate existing labeling errors. Our experimental results suggest that the benefits of HLP generally outperform these limitations, particularly for smaller architectures. Future research should focus on developing robust HLP techniques for noisy scenarios and exploring synergies with model compression techniques.

In conclusion, by enhancing dataset quality and boosting model performance, particularly for smaller models, HLP presents a promising avenue for advancing the field. Our findings indicate that HLP could serve as a useful technique for datasets with hierarchical label structures, particularly during dataset curation, training, or evaluation phases.

\section*{Acknowledgment}

This work was granted access to the HPC resources of IDRIS under the allocation AD011014754 made by GENCI. Support from the ANR-3IA Artificial and Natural Intelligence Toulouse Institute ANITI (ANR-19-PI3A-0004) is gratefully acknowledged.

\IEEEtriggeratref{15}

\label{sec:refs}

\end{document}